\documentstyle[]{l-school}
\def\be{\begin{equation}}
\def\ee{\end{equation}}
\newcommand {\bea} {\begin{eqnarray}}
\newcommand {\eea} {\end{eqnarray}}
\def\beq{\begin{equation}}
\def\eeq{\end{equation}}
\def\bear{\begin{eqnarray}}
\def\eear{\end{eqnarray}}
\newcommand {\eqref} [1] {(\ref {#1})}

 \def\ra{\rightarrow}

\def\lnc#1{{\it Lett. Nuovo Cim.} {\bf #1}}
\def\pl#1{{\it Phys. Lett.} {\bf #1B}}

\def\prd#1{{\it Phys. Rev.} {\bf D#1}}

\def\np#1{{\it Nucl. Phys.} {\bf B#1}}

\def\jmath#1{{\it J. Math. Phys.} {\bf #1}}

\def\dslash{\not{\hbox{\kern-2pt $\partial$}}}
\def\Dslash{\not{\hbox{\kern-2pt $D$}}}
\def\Qslash{\vert{\hbox{\kern-5pt Q}}}
\def\Rslash{\vert{\hbox{\kern-5.5pt R}}}
\def\hslash{\not{\hbox{\kern+1.5pt h}}}

\def\({\left\lbrack}           \def\){\right\rbrack}
\def\{{\left\lbrace}           \def\}{\right\rbrace}

\def\pa{\partial}

\def \pa{\partial}

 \def\pa{\partial}
\def\bpa{\bar\partial}

\def\Q{$QCD_2$\ }

\def\pl#1{{\it Phys. Lett.} {\bf #1B}}

\def\prd#1{{\it Phys. Rev.} {\bf D#1}}

\def\np#1{{\it Nucl. Phys.} {\bf B#1}}

\def\jmath#1{{\it J. Math. Phys.} {\bf #1}}

\def\nc{N_c}
\def\nf{N_f}
\def \j{h^{-1}\pa h }
\def \bj{ h\bpa h^{-1}}

\def\boxit#1{\vbox{\hrule\hbox{\vrule\kern3pt
\vbox{\kern3pt#1\kern3pt}\kern3pt\vrule}\hrule}}

\def\half{{1\over 2}}
\begin{document}
\markboth{J.Sonnenschein}{SCREENING and CONFINEMENT in $QCD_2$}

\title{More on Screening and Confinements in 2D $QCD$}
\author{J.Sonnenschein}
\institute{ School of Physics and Astronomy,\\
 Beverly and Raymond-Sackler Faculty of Exact Sciences,\\
 Tel-Aviv University, Ramat-Aviv, Tel-Aviv 69978, Israel}
\maketitle

\section{Introduction}

Gross et al. \cite{Gross} have argued that massless  dynamical
fermions {\bf screen}
  heavy external charges,
 even if they  are in a
representation which  has zero 
``$N_c$-ality" ( $Z_{N_c}\sim 0$), 
namely,  even if the dynamical quarks cannot compose the external
ones.
Confinement is restored  as soon as the dynamical
fermions get some non-trivial mass (unless they can compose the
representation of the external quarks). 
This conjecture was  ``proven" in \cite{Gross} for the following systems
(i) abelian case, (ii) $SU(N_c=2)$ with the dynamical fermions in
{\bf 3},
(iii)  spinor {\bf 8} of $SO(8)$.

In the present talk I will describe two papers \cite{FSsol,ASnf} 
where we presented 
 further evidence  for screening of massless  
 fundamental and adjoint fermions of any non-abelian  group.
 In those works we have  used the following methods:
(i) Deducing the potential from  solutions of   
   the equations
of motion of the  bosonized action.
(ii)  Bosonizing   also the heavy external
charges. In this framework 
{\bf Confinement} is manifested via the 
 absence of soliton solutions
of  unbounded quarks, and  {\bf Screening} 
 implies finite-energy quark static solutions 
(iii) Large $N_f$ analysis of the system including  analyzing also
  the next to the leading order behavior. 

 \subsection{ background remarks}

(1) The screening mechanism in the massless Schwinger model could be
attributed  to the 
 {\bf exchange of  the 
``massive photon"}. 
The non-abelian counterpart is
clearly  much more complicated.
However,  in the limit of
large number of flavors $N_f$ (with finite $N_c$),
the non abelian theory resembles 
{\bf a collection of
$N_c^2-1$ abelian theories.}
In that limit the spectrum includes Schwinger like
$N_c^2-1$ massive modes that induce screening.

(2) The ``massive gauge states"  are an indication of the
``non-confining" structure of the spectrum. 
Even though they are gauge invariant states,  they
are in the adjoint rep. of a ``global
color symmetry". 
These states
  had  already  been
pointed  out in an earlier work,\cite{FHS}
based on a BRST analysis and a special parametrization of the
gauge configurations.
However, in that paper  we were not able
to rigorusly show that  indeed they were part of the BRST cohomology.
Note, that even if they are not in space of physical states, 
{\bf the
massive states could nevetheless be responsible for the
screening potential.}

(3) When passing from a screening picture at large $N_f$ and finite
$N_c$ to the domain of a small  $N_f$, one can anticipate
two types of  scenarios: 
(i) A {\bf  smooth transition}, 
namely, the screening behavior persists 
all the way down to $N_f=1$.
(ii) A {\bf ``phase transition" }
 at a certain value of 
$N_f$ and
 confining nature below it.

(4) Are the 
 massive modes of large $N_f$  an artifact
of the abelianization of the theory?
To check this possibility we have searched for 
{\bf non-abelian solutions}
of the equations of motions. 
Indeed, we found new non-abelian  gauge solutions
that are also  massive.\cite{FSsol} 
We believe that this presents 
certain  evidence in favour of  option (i).

(5) Low lying baryonic states, were 
determined in the semiclassical picture of the strong coupling
limit.\cite{DFS} 
Does it contradict the
screening  nature of the theory?
It does not, 
since the baryons were discovered only for massive quarks and not for
massless ones. 

\subsection{ The outline of the talk}

(1)  Review of Bosonization in $QCD_2$

(2)  Equations of motion  of \Q in the presence 
of external
currents

(3) Solutions of the equations without external 
quarks

(4) Solutions of the equations with external 
current and the
potential

(5)  Bosonized external currents

(6)  Large $N_f$ expansion

(7)  $N=1$ Super Yang-Mills

\section{ Review of Bosonization in $QCD_2$ }

\subsection{  Dirac fermions in the fundamental representation.}
The action of  massive \Q \cite{FS}

\bear
S_{QCD_2}&=&S_1(u)-{1\over
2\pi} \int d^2 z Tr(iu^{-1}\pa u \bar A \cr
&+& iu\bar \pa
u^{-1} A + \bar A u^{-1}A u-A \bar A )\cr
&+&{m^2\over{2\pi}}\int d^2z :Tr_G[u + u^{-1}]:
+{1\over e_c^2}\int d^2 zTr_H[F^2]\cr
\label{mishwzw}
\eear 
where $u\in U(N_f\times N_C)$;
$S_k(u)$ is a level $k$ WZW model;
$A$ and
$\bar A$  $\in$ algebra of $H\equiv SU(N_C)$;
$F=\bar\pa A-\pa\bar A+i[A,\bar A]$;
and where  $m^2= m_q\mu {1\over2}e^\gamma$,
where
$\mu$ is the normal ordering mass.

 Notice that the  space-time has a
Minkowski signature, 
and we use the following notations 
 $B\equiv
B_+$ and $\bar B\equiv B_-$.

The action for massless fermions can be simplified 
\bear S &=&S_{N_f}(h) -{\nf\over
2\pi} \int d^2 z Tr(ih^{-1}\pa h \bar A \cr
&+& ih\bar \pa
h^{-1} A + \bar A h^{-1}A h-A \bar A )\cr
&+&S_{N_C}(g)+{1\over2\pi}\int d^2z[\pa\phi\bpa\phi]
{1\over e_c^2}\int d^2 zTr_H[F^2]\cr
\label{mishwzwghl}
\eear
using the following
parametrization  $u \equiv ghle^{i\sqrt{4 \pi\over N_CN_f}\phi}$
where $h\in SU(\nc)$, $g\in SU(\nf)$, $l\in {U(\nc\nf)\over [SU(\nc)]_{\nf}
\times
[SU(\nf)]_{\nc}\times U_B(1)}$ and setting $l=1$. 
\subsection{ Majorana fermions in the adjoint representation.}

An action of bosonized Majorana fermions in the
adjoint representation\cite{FSsol}

\bear
 \label{mishwzwad}
S &=&\half S(h_{ad}) -{1\over
4\pi} \int d^2 z Tr(ih_{ad}^{-1}\pa h_{ad} \bar A + iu\bar \pa
h_{ad}^{-1} A \cr
&+& \bar A h_{ad}^{-1}A h_{ad}-A \bar A )
{m^2\over{2\pi}}\int d^2z\ Tr_G:[h_{ad}
+h_{ad}^{-1} ]:
{1\over e_c^2}\int d^2 zTr_H[F^2]\cr
\eear
where  $h_{ad}$ are
$(\nc^2-1)\times (\nc^2-1)$ matrices; the  
 Virasoro anomaly $c_{vir}=\half (\nc^2-1)$,
 the affine Lie algebra  anomaly  $=\nc$
 the conformal dimension  $\Delta h_{ad}=\half$ 
and the factor $\half$ in $\half S(h_{ad})$ comes from the
reality of the Majorana fermions.

\section{ Equations of motion  of \Q in the presence of external currents}

Studying  the  quantum system by  analyzing the corresponding equations of
motion is a justified approximation only provided that the 
{\bf classical
configurations 
dominate the functional integral.}
Such a scenario can be
achieved in the limit of a
large number of flavors. 
where ${1\over
N_f}$ plays the role of $\hslash$. 

The colored sectors  of the action in ($A=0$ gauge)
takes the form
$$S =N_f\{S_{1}(h) -{1\over
2\pi} \int d^2 z Tr(ih^{-1}\pa h \bar A )
+{1\over \tilde e_c^2}\int d^2 zTr[\pa \bar A]^2\} $$

where $\tilde e_c= e_c\sqrt {N_f}$.

The equations of motion which follow from the variation of the
action \label{mishwzwghl}   with respect to $h$  are
\bear
&\bpa(\j) + i\pa \bar A +i[\j,\bar A] =0\cr
&\pa(\bj)  -i\pa (h\bar A h^{-1}) =0\cr
\eear
A similar  result holds for  $ h_{ad}$.
External  currents are coupled to the system by
adding to the action \eqref{mishwzwghl} or \eqref{mishwzwad}  
$${\cal L}_{ext}={1\over
2\pi} \int d^2 z Tr(J_{ext} \bar A + \bar J_{ext}  A).$$

Now the equations of motion take the form
\bear
&\pa^2 \bar A  + \alpha_c(iN_f\j+J_{ext}) =0\cr
&\pa\bpa \bar A +[i\pa \bar A, \bar A] - \alpha_c[N_f(i\bj+h \bar A h^{-1}
-\bar A)+\bar J_{ext}]
=0\cr\label{misheomA}
\eear
where $\alpha_c={e_c^2\over 4\pi}$.
It follows from the equations of motion \eqref{misheom}\ and
\eqref{misheomA}\ that  both the dynamical currents $j_{dy}={iN_f\over 2\pi}\j,\
\bar j_{dy}={iN_f\over 2\pi}[\bj -ih \bar A h^{-1} +i\bar A]$ as well as the
external currents are covariantly conserved, which for
$A=0$ reads 
\be
\bar D j_{dy} +\pa \bar j_{dy}= 0\qquad
\bar D J_{ext} +\pa \bar J_{ext}= 0.\label{mishccc}
\ee
with $\bar D= \bpa - i[\bar A , ]$.

One can  eliminate the dynamical current, and then one finds
\be
\pa\bpa \bar A +[i\pa \bar A, \bar A] +
\alpha_c(N_f\bar A-\bar J_{ext})=0 \label{misheqA}\ee
In fact the equation
 one gets   $\pa [l.h.s]\eqref{misheqA}=0$.  
The antiholomorphic function can be eliminated
by  fixing  the residual gauge
invariance  
$\bar A\rightarrow iu^{-1}\bpa u + u^{-1}\bar A u$, 
with $\pa u=0$.


\section{ Solutions of the equations without external quarks}

We enlist several statements about the solutions:

(1)The equation of motion is
not invariant but rather covariant with respect to the ``global
color" transformation 
$A\rightarrow  u^{-1}A u$ with a constant $u$.
So any solution is in the adjoint rep. of the ``global
color" group. 

(2) {\bf ``abelian"  massive mode} 

Consider a configuration of the form
$\bar A\equiv T^a \bar A^a(z,\bar z)= T^a\delta^{a,a_0} \bar {\cal
A}(z,\bar z)$
then the commutator term
vanishes and  ${\cal A}$ has to solve

$\pa\bpa \bar {\cal A} + \tilde\alpha_c\bar{\cal A}=0$
with $\tilde\alpha_c= N_f\alpha_c$.

(3) {\bf No soliton solutions}
Let us now check whether the equations admit soliton solutions.
For static configurations 
$$\pa_1^2\bar A -\sqrt{2}[i\pa_1 \bar A, \bar A] -2 \tilde\alpha_c\bar A=0
\label{misheqAs}$$
Multiplying the equation by $\bar A$, taking the trace of the result and
integrating over $dx$ one finds after a partial integration that
$\int dx[ Tr [(\pa \bar A)^2 + 2\tilde\alpha_c A^2]=0$
which can be satisfied only for a vanishing $\bar A$.

(4) {\bf Abelian Bessel function}
Consider an ansatz for the solution of the form
$\bar A = zC(z\bar z)$
and $C$ is determined
by the equation 

$ \rho C" + 2C' + \tilde\alpha_c C =0$, where $C'=\pa_{z\bar z} C$.

The solution for the gauge field 
$$ \bar A = {\bar A_0z\over \sqrt{ \alpha_c z\bar z}} J_1(2\sqrt { \alpha_c
z\bar
z})\label{mishBes}$$
where $\bar A_0$ is an arbitrary constant matrix.

(3) {\bf Non-abelian solution}

Consider in  the special case of $SU(2)$  the configuarion 
$\bar A = e^{-i\theta \tau_0}\bar A_0 e^{i\theta
\tau_0}$ with  a constant matrix 
$\bar A_0=e_0\tau_0+\bar e\tau+e\bar\tau$
 Plugging this ansatz into eqn.
\eqref{misheqA}\ with no external source
 one finds that there is a solution provided that 
 $\theta= \theta_0 +k\bar z+\bar k z$

where $k$,
  $\bar k$ and $\theta_0$ are constants.
Indeed the following gauge field
$$\bar A ={\tilde\alpha_c-k\bar k\over \bar k}\tau_0 +
\sqrt{(k\bar k-\tilde\alpha_c)\tilde\alpha_c\over 4\bar k^2}[e^{-i\theta}
\tau+e^{i\theta} \bar \tau]$$ 
is a `` non-abelian solution". 
Setting $\theta_0=0$ requires that $(k\bar
k-\tilde\alpha_c) >0$.
The solution  is {\bf truly non-abelian}.
The corresponding $F$ is $F=
 i\bar k e^{-i\theta \tau_0}(\bar e\tau-e\bar\tau)e^{i\theta \tau_0}$.
Performing a gauge transforamtion with 
$U=e^{-i\theta \tau_0}$ 
$\ra$ $F_U= i\bar k(\bar e\tau-e\bar\tau)$, $\bar A_U =\bar
A_0+k\tau_0$  $A_U=\bar k\tau_0$. 
$A_U,\bar A_U$ and $F_U$ are constants and no
two commuting. 
Furthermore an abelian gauge configuration of the
form $\bar A=-i\bar kz(\bar e\tau-e\bar\tau)$ and $A=0$ that leads
to the same $F$ is not connected to $A_U,\bar A_U$ by a gauge
transformation.

\section{The energy-momentum tensor and the spectrum
of non-abelian solution}

First  we have   to compute the  components of the energy momentum
tensor 
$T\equiv
T_{zz}, \bar T\equiv T_{\bar z\bar z}, T_{\bar z z}$  
that corresponds to
the action \eqref{mishwzwghl}.

Only the colored part of the energy momentum tensor is relevant to
our discussion.
 
$$T={\pi\over N_f+N_c}:Tr[j_{dy}j_{dy}]:\ \
\ \ T_{\bar z z}= {1\over 8\pi\alpha_c}
 Tr[(\pa\bar A)^2]
\label{mishemt}$$
 where the   dynamical currents  which
were defined below eqn.\eqref{misheomA}\ are
\bear
j_{dy}=&{iN_f\over 2\pi}h^{-1}\pa h= -{1\over 2\pi\alpha_c}\pa^2\bar A;\ \ \
\cr
  \bar j_{dy}=&{1\over 2\pi\alpha_c}(\pa\bpa\bar A +i[\pa  \bar A,\bar A])=
  -{iN_f\over 2\pi}h^{-1}\bpa h=-{N_f\over 2\pi}\bar A 
\label{msihJdy}
\eear
 It is natural in the light cone gauge to use a 
{\bf light
front quantization.}
We take $ z$ to denote the
space coordinate. 
$\bar P$ and $ P$ are integrals over $T$ and $T_{\bar z z}$.
 The  masses of the states $=$  the eigenvalues of
$M^2=P\bar P$.

A proper normalization of the fields is introduced and discussed
in \cite{FSsol}. Using this normalization we find the following
 expectation values of the energy momentum tensor 
\bear
<T>&={1\over 8\pi}{N_f^2\over N_f+N_c} (\bar k)^2
 { k\bar k\over \tilde\alpha_c}(1-{\tilde\alpha_c\over k\bar k})\cr
<T_{\bar z z}>&={N_f\over 16\pi} (k\bar k) 
 (1-{\tilde\alpha_c\over k\bar k})
\label{mishemtt}
\eear
The corresponding momenta

$$ \bar P={1\over 8\pi}{N_f^2\over N_f+N_c}( L \bar k) \bar k
 { k\bar k\over \tilde\alpha_c}(1-{\tilde\alpha_c\over k\bar k})\ \ \
\  P={N_f\over 16\pi}( L \bar k)  k
 (1-{\tilde\alpha_c\over k\bar k})$$

  The non-abelian state is
characterized by  masses

$$M^2={n^2\over 32}{N_f^3\over N_f+N_c}
 (k\bar k){ k\bar k\over \tilde\alpha_c}
(1-{\tilde\alpha_c\over k\bar k})^2$$

Note that 

(1)  $(k\bar k-\tilde\alpha_c)>0$\ \ \  
(the case of zero $k\bar k-\tilde\alpha_c$ 
corresponds to vanishing $\bar A$).

(2) $M$  starts from $M=0$  and
grows up linearly in $k\bar k$ for $k\bar k>>\tilde\alpha_c$.

(3) the solution is singular for $\tilde \alpha_c=0$.

\section{ Solutions of the equations with external current}

 We turn now  on a  covariantly conserved (eqn. \eqref{mishccc})
 external current $J_{ext}$ 

(1) Abelian solutions of the equations of motion
are easily constructed.

 For instance for a pair of quark
anti-quark as an external classical  source
$\bar J^a{ext}=T^a\delta^{a,a_0}Q[\delta(x_1-R)-\delta(x_1+R)]$
the abelian solution is
$$ \bar A=\half\sqrt{2\alpha_c}T^a\delta^{a,a_0}
Q[e^{-\sqrt{2\tilde\alpha_c}|x_1-R|}-
e^{-\sqrt{2\tilde\alpha_c}|x_1+R|}].\label{mishabexso}$$

Inserting this expression into ${1\over 2\pi}\int dx_1Tr[A\bar
J_{ext}]$ 

one finds the usual {\bf screening potential} 

$$V(r)= {1\over
2\pi}\sqrt{2\alpha_c}Q^2(1-e^{-2\sqrt{2\alpha_c}|R|})Tr[(T^{a_0})^2]
\label{mishVR}$$.

(2){\bf ``non-abelian" solutions} 

The  $SU(2)$ ``non-abelian solution"  is a solution also in
case of a constant external current  
$\bar J=\tau^a\delta^{a,0}J_0$ 
with the trivial modification that
the coefficient of 
$[e^{-i\theta}
\tau+e^{i\theta} \bar \tau]$ is now
$\sqrt{\tilde\alpha_c [{J_0\bar k\over
N_f}+(k\bar k-\tilde\alpha_c)]\over 4\bar
k^2}$.

 Consider now an external current of the form
$\bar J= \bar J_0(\bar z) \tau_0 $.
A solution in that case is
$$\bar A=(f_0+J_0(\bar z))\tau_0 +
[g(e^{[-i(k\bar z + \bar
k z+I)]})\bar \tau+c.c]$$
where $\bpa I(\bar z)={1\over N_f}\bar J_0(\bar z)$ with
 $f_0$ and $g$  related to $k$ and $\bar k$ as given in eqn.
\cite{mishsolu}. 
In the case of light-front ``static" current
$\bar J_{ext}(\bar z)$ 
 In particular 
a quark anti-quark pair 
$\bar J^a_{ext}=\half\tau^a\delta^{a,a_0}Q[\delta(\bar z-R)-\delta(\bar z+R)]$
The corresponding solution has 
$\epsilon(\bar z-R)$
and  $\epsilon(\bar z+R)$ factors in $\theta$.
The corresponding potential is a constant thus  non-cofining.

\section{ Bosonized external currents}

Another approach to the coupling of the dynamical fermions to external
currents is to 
{\bf bosonize the ``external" currents.}

Let us briefly summarize first the abelian case.
Consider external
fermions of mass M and charge $qe$ 
described by the real scalar filed $\Phi$ and
dynamical fermions of unit charge $e$ and mass $m$ described by
 the scalar $\phi$. 

 The Lagrangian of the combined system after
integrating  out the gauge fields is given by
\bear
{\cal L}&=& \half(\pa_\mu\phi\pa^\mu\phi) +
m\Sigma[cos(2\sqrt{\pi}\phi)-1]
+ \half(\pa_\mu\Phi\pa^\mu\Phi)\cr 
&+& M\Sigma[cos(2\sqrt{\pi}\Phi)-1]
-{e^2\over 2\pi}(\phi + q\Phi)^2\cr
\label{mishbosex}
\eear

Let us look for
static solutions of the corresponding equations of motion with
 finite energy. 
Take, without loss of generality,
 $\phi(-\infty)=\Phi(-\infty)=0$.
From the M term we get
$\Phi(\infty)=\sqrt{\pi}N$ with $N$ integer.  
For $m\ne 0$ we also get
$\phi(\infty)=\sqrt{\pi}n$. Now from the $e^2$ term,  $n+qN=0$.
Thus, for instance for    $N=1$ finite energy solutions occur only for
$q=-n$. 
For $m=0$ only $\Phi(\infty)=\sqrt{\pi}N$
and then  a finite energy solution for $N=1$ is if
 $\phi(\infty)=-\sqrt{\pi}q$.
So, when $q=-n$, $\ra$ the screening phase,
$q\ne -n$ 
for $m\ne 0$  $ \ra$ confinement phase 
\hskip 1 cm 
$m=0$\ $\ra$ screening .

Proceeding now to the QCD case  one can
consider  several different possibilities 
$(J_{ext}^{F},j_{dy}^{F}),
(J_{ext}^{F},j_{dy}^{ad}),(J_{ext}^{ad},j_{dy}^{F}),(J_{ext}^{ad},j_{dy}^{ad})$
and with dynamical fermions that can be either massless or massive.

The system of dynamical adjoint fermions and external fundamental quarks can
be described by an action which is the sum of  \eqref{mishwzwad}\
 and \eqref{mishwzw}.

Integrating over the gauge degrees of freedom one is left with the
terms in \eqref{mishwzw}\ and  \eqref{mishwzwad}\ that do not include coupling to gauge field
together with a current-current non-local interaction term.

For the interesting case of dynamical quarks in the adjoint and external in
fundamental we get for the interaction term
$$\int d^2 z \sum_a\{{1\over \pa}[Tr(T^a_Fi u_{ext}^{-1}\pa u_{ext})+
{i\over 2} Tr(T^a_{ad}\j )]\}^2\label{mishscr}$$
where $u$ is defined in \eqref{mishwzw},
$T^a_F$ are the  $SU(\nc)$ generators expressed
as $(\nc\nf)\times(\nc\nf)$ matrices in the fundamental rep.
of $U(\nc\times \nf)$
$T^a_{ad}$ the
$(\nc^2-1)\times (\nc^2-1)$  matrices
in the $SU(\nc)$ adjoint rep. 
For simplicity we discuss from here on the case of a single
flavor.

Let us first  consider the
case of external adjoint  
quarks $u_{ext}(x)\in SU(\nc)\times U_B(1)$,
 
$$u_{ext}\,=\,\pmatrix{\matrix{e^{-i\Phi}&&&&\cr
&e^{i\Phi}&&&\cr &&1&&\cr &&&\ddots&\cr&&&&1\cr}}\label{mishstso} $$
($\Phi$ is not normalized canonically here). 

The reason that we take a diagonal ansatz is that it corresponds,
 to
a minimal energy configuration.

Ansatz \eqref{mishstso}\ corresponds to $Q^1_{ext}\bar Q^2_{ext}$, namely,
to an external adjoint state. We expect this state to be screened
by the adjoint dynamical fermions. With this ansatz  ${1\over
\pa}(ih_{ext}^{-1}\pa h_{ext})$ takes the form
$$\pmatrix{\matrix{\Phi&&&&\cr &-\Phi&&&\cr &&0&&\cr
&&&\ddots&\cr&&&&0\cr}}\label{mishpastsoo} $$ It is thus clear that
only $T^3_F$ contributes to the trace in \eqref{mishscr}. To show the
dynamical configuration that screens, take 
$$log\ 
h_{ad}=\pmatrix{\matrix{0&-\phi&0&&\cr \phi&\  0&0&&\cr
0&0&0&&&\cr &&&&\ddots&\cr&&&&&0\cr}}\label{mishpastsh} $$ with
the matrix that contributes  to the $Tr$ in \eqref{mishscr},
$$ T_{adj} = i\pmatrix{\matrix{\ \ 0& 1& 0&&&\cr
-1& 0& 0 &&&\cr\ \  0& 0& 0&&&\cr&&&0&&\cr
&&&&\ddots&\cr&&&&&0\cr}}\label{mishpastso1} $$ 
which corresponds to the generator of rotation in direction 3
for the sub $O(3)$ of first three indices, 
thus obtaining the a term
proportional to  $(\Phi+\phi)^2$  emerging from \eqref{mishscr}. The mass terms for
$u_{ext}$ and $h$ are now proportional to  $(1-cos\Phi)$ and 
$(1-cos\phi)$ respectively.  A boundary condition
$\Phi(\infty)=2\pi$ can be cancelled in the interaction term
 by the boundary condition $\phi(\infty)=-2\pi$.

Let us examine now the case of a single external quark
$$u_{ext}\,=\,\pmatrix{\matrix{e^{-i\Phi}&&&\cr
 &1&&\cr &&\ddots&\cr&&&1\cr}}\label{mishstso} $$
Its  contribution  to  the  interaction term
is $\sum_{i=1}^{\nc-1}(\eta_i\Phi+``dyn")^2$ where
$\eta_i=
{1\over
\sqrt{2i(i+1)}}$ and $``dyn"$ is the part of the dynamical quarks.

If again we take a configuration of the dynamical quarks based
on a single scalar like in \eqref{mishpastsh}\
we get altogether an $e^2$ term of the form
$(\half\Phi+\phi)^2$.  

Now if $\Phi(\infty)=2\pi$ one cannot find a finite
energy solution since from the mass term  $\phi(\infty)=2\pi n$,
and thus there is no way to cancell the interaction term. 

If,
however, we consider$m=0$ there is no
constraint on   $\phi(\infty)$ so it can be taken to be equal
$-\pi$, and thus  again a screening situation is achieved.

This argument should be supplemented  by showing that one cannot
find another configuration besides \eqref{mishpastsoh}\  that may cancel
 the $\eta_i\Phi$ term in $(\eta_i\Phi+``dyn")^2$, for
$SU(N_c)$ with $N_c\geq 3$. 
\section{Large $N_f$ expansion}

Consider a system of a quark in the fundamental representation placed at a
distance  of $2R$ from an anti-quark that transforms in the anti-fundamental
representation. This can be expressed as the following classical c-number
charge density
\beq
  \label{rho} \rho ^a=\delta ^{a1} (\delta (x-R) - \delta (x+R))
\eeq
Strictly speaking, one is allowed to introduce classical charges
and neglect  quantum fluctuations  only if the external charges
 transform in  a large color representation.
(This is an analog of
the statement that only for quantities of large angular  momentum quantum
fluctuations are suppressed.)
Moreover, by choosing \eqref{rho}, there is an obvious   ``abelian''
self-consistent solution of the equations \eqref{set}for which
  all the dynamical quantities points in the '1' direction and thus all the
 commutators vanish.

One way to overcome these obstacles is to search for ``truly" non-abelian
solutions of the equations,as was discussed in section 4.
Here we proceed by  implementing   Adler's\cite{adler}
semi-classical approach  for introducing  static external quark
charges.
 In this
approach the quarks color charges satisfy non-abelian $SU(N_c)$ color algebra
so that  the external quark charge density  takes the form
\beq
 \label{rho2} \rho ^a = Q^a \delta (x-R) + \bar Q^a \delta (x+R)
\eeq
 $Q^a$ and $\bar Q^a$ are
  in   $({\bf N_c}, 1)$ and   $(1,{\bf\bar  N_c})$ representations of
 $SU(N_C) \otimes SU(N_C)$ group respectively.
The algebra of those operators  was worked out in \cite{adler,giles}.
is reviewed briefly in the appendix.

The expansion in  ${1\over N_f}$
is defined as usual by taking $N_f\ra \infty$ while
 $\tilde\alpha_c= {{e^2N_f}\over4 \pi}$ is kept fixed. 
One can solve the equations  iteratively as follows.
 A   solution for  $\bar A$,  $J$ and $J_{ext}$
of the equations  expanded to a given  order in $e$
 is inserted back to the equations as a source to
determine the next order solution.
A similar treatment in four dimensions is given
 in refs.\cite{jackiw,arodz}.

 The formal expansion in $e$ is as follows
\bea
 && \bar A=e  A^{(1)} + e^3  A^{(3)} + e^5 A^{(5)} + \dots \nonumber \\
 && J = J^{(0)}+ e^2 J^{(2)}+ e^4 J^{(4)} +\dots \label{expansion} \\
 && J_{ext}= e^2 j^{(2)}+ e^4 j^{(4)}+ e^6 j^{(6)}+ \dots \nonumber
\eea

After performing the interation one finds that the potential
up to next to leading order takes the form

\bea
  \lefteqn{V(2R)=
 \mu {\pi \over {2N_f}} {{N_c^2-1}\over {2N_c}} (1-e^{-\sqrt{\tilde\alpha}2R}) }
\label{potential2} \\ &&
 +\mu {({\pi \over {2N_f}})}^3 {{N_c^2}\over 2}{{N_c^2-1}\over {2N_c}} \left (
   {(1-e^{-\sqrt{\tilde\alpha}2R})} ^2 
- \sqrt{\tilde\alpha}2R \ e^{-\sqrt{\tilde\alpha}2R} (1-e^{- \sqrt{\tilde\alpha}2R})
 \right )
  \nonumber
\eea

Thus, the potential that includes
the first correction to the abelian one approaches a constant value
at large distances where the force between the  external quark and the
anti-quark vanishes.

\section{Massive $QCD_2$}
 The bosonized action
 of massive $QCD_2$ with $N_f$ fundamental representations
was given in \eqref{mishwzw}. 
We can now expand the mass term in a ( non-local) power series in $J$
 as follows
$$ h=1-{2\pi\over iN_f}{1\over \pa J} +{2\pi\over iN_f}^2{1\over \pa
  J^2} -{1\over \pa}{1\over \pa J}J+ ...$$
 
Now the solution of the equation in the presence of \eqref{rho2} is
\beq
\bar A = -{e\over \sqrt 2}(1+ {e ^2 N_f^2\over 8\pi ^2 m^2})^{-1}(Q \mid x-R
\mid +\bar Q \mid x+R \mid)
\eeq
Substituting $\bar A$ in the potential yields,
\beq
 V=- {e^2\over 2} (1+ {{e^2 N_f^2}\over{8\pi^2 m^2}})^{-1}Q\bar Q \times 2R
={\mu^2 \pi \over 2 N_f} (1+ {\sqrt{\tilde\alpha}\over 8\pi Cm_q} )^{-1}C_2({\cal R})
\times 2R
\eeq

The same expression for the potential in the abelian case was
obtained in \cite{Gross}.
Thus the dominant ${1\over N_f}$ contribution exhibits a
confinement behavior.
It should be emphasized that in the above analysis it is assumed
that the external charges cannot be composed by the dynamical ones.

\section{Supersymmetric Yang-Mills}
The supersymmetric YM (SYM)  action \cite{ferrara} is the following
\beq
 \label{sym} S = \int d^2 x \ tr \left ( -{1\over 4} F^2_{\mu\nu} +  i\bar
 \lambda\Dslash\lambda
 +{1\over 2} (D_\mu \phi )^2 + 2ie \phi \bar \lambda \gamma _5 \lambda
\right ) ,
\eeq
The gluon equation of motion  now reads
\beq
\label{gluon}
D_{\mu} D^{\mu} F = e \epsilon ^{\mu \nu} D_{\mu} (J^\lambda_\nu +
J^\phi _\nu),
\eeq
where $J^\lambda_\mu$ denotes the gluino vector current and $J^\phi
_\mu$ denotes the scalar vector current.
 The equation for  the divergence of the
 fermionic axial current.
\beq
 \epsilon ^ {\mu \nu} D_\mu J^\lambda _\nu = -{eN_c \over \pi} F + ie
(\phi \bar \lambda \lambda + \bar \lambda \lambda \phi)
\eeq
The factor $N_c$ which appear in the anomaly term is due to
the
 adjoint gluinos which  run in the
anomaly loop. A similar equation holds for the scalar current
\beq
 \epsilon ^{\mu \nu} D_\mu J^\phi _\nu = -{eN_c \over {2\pi}} F -
 \epsilon ^{\mu \nu} \partial _\mu (-2i[\phi ,\partial _\nu \phi] +
 2e[\phi,[A_\nu,\phi]])
\eeq
This equation was obtained by applying the point splitting technique
to the scalar current.

Thus the quantum version of equation \eqref{gluon} is
\beq
(D_{\mu} D^{\mu} +{3e^2 N_c \over {2\pi}})F =  ie^2  (\phi \bar
 \lambda \lambda  + \bar \lambda \lambda \phi) -e \epsilon ^{\mu \nu}
 \partial_\mu
(-2i[\phi ,\partial _\nu \phi] + 2e[\phi,[A_\nu,\phi]])
\eeq
 which means that the gluon propagator has only a single pole which is
massive.

The implication of the  last equation on the potential between external
charges is clear. The interaction mediated by the exchange of these  massive
modes is necessarily a screening one.
The potential  takes the form of eqn.\eqref{potential2} with a range that
behaves like $\sim [e\sqrt{N_c}]^{-1}$.

\end{document}